\documentclass[twocolumn,showpacs,amsfonts,aps,prc,floatfix,superscriptaddress]{revtex4-1}

\usepackage[colorlinks=true, pdfstartview=FitV, linkcolor=red, citecolor=blue, urlcolor=blue]{hyperref}
\usepackage{amsmath}
\usepackage{bm}
\usepackage{graphicx}

\begin{document}

\title{
Effective distributions of quasi-particles for thermal photons
}

\author{Akihiko Monnai}
\email[]{amonnai@riken.jp}
\affiliation{RIKEN BNL Research Center, Brookhaven National Laboratory, Upton, NY 11973, USA}
\date{\today}

\begin{abstract}
It has been found in recent heavy-ion experiments that the second and the third flow harmonics of direct photons are larger than most theoretical predictions. In this study, I construct effective parton phase-space distributions with in-medium interaction using quasi-particle models so that they are consistent with a lattice QCD equation of state. Then I investigate their effects on thermal photons using a hydrodynamic model. Numerical results indicate that elliptic flow and transverse momentum spectra are modified by the corrections to Fermi-Dirac and Bose-Einstein distributions.
\end{abstract}


\maketitle

\section{Introduction}
\label{sec1}
\vspace*{-2mm}

The discovery of collective dynamics of the QCD media created in BNL Relativistic Heavy Ion Collider (RHIC) and CERN Large Hadron Collider (LHC) implies early equilibration of low-momentum gluons and light quarks, and realization of a quark-gluon plasma (QGP) fluid \cite{RHIC:summary, LHC:v2}. 
Dynamical description of a QCD system in the vicinity of the quark-hadron crossover is generally a difficult issue. The fluidity of the system, on the other hand, is believed to allow an effective description of the bulk dynamics of heavy-ion collisions \cite{Schenke:2010rr} where aside from the initial conditions the information of QCD comes into the formalism through the equation of state (EoS), the transport coefficients and the chemical reaction rate. Particle spectra and flow harmonics \cite{Ollitrault:1992bk, Poskanzer:1998yz} of hadrons are reasonably well-explained in the framework of hydrodynamics with small viscosity. Direct photons, on the other hand, are found to be non-trivial, because elliptic flow $v_2$ \cite{Adare:2011zr,Lohner:2012ct} and triangular flow $v_3$ \cite{Mizuno:2014via} of direct photons undershoot those measured in the experiments by a factor when estimated in conventional models. Here direct photons are defined as the photons which does not originate from hadronic decay. Initial hard photons, named prompt photons, do not have azimuthal anisotropy, but medium-induced soft photons, known as thermal photons, can inherit anisotropy from the medium. The apparent discrepancy poses theoretical challenges to the community \cite{Chatterjee:2005de, Chatterjee:2008tp, Chatterjee:2011dw, Holopainen:2011pd, Chatterjee:2013naa, Chatterjee:2014nta, Liu:2009kta, vanHees:2011vb, Dion:2011pp, Basar:2012bp, Bzdak:2012fr, Goloviznin:2012dy, Hattori:2012je, Liu:2012ax, Linnyk:2013hta,Muller:2013ila,Monnai:2014kqa,McLerran:2014hza,Monnai:2014taa,vanHees:2014ida,Shen:2013cca, Gale:2014dfa,Biro:2015iua}.

Despite the experimental evidences that the system is likely to be strongly-coupled, heavy-ion models often employ the phase-space distributions of ideal gases. In general the equilibrium phase-space distributions in an interacting system should be subject to non-ideal corrections. They are known to play important roles in phenomenology in non-relativistic systems -- carbon dioxide, for example, has a solid phase due to van der Waals force which arises from interaction corrections. It should be noted the deviations from the ideal gas picture is a different concept than viscous corrections because the system is in local equilibrium with no entropy production. 

In heavy-ion collisions, hydrodynamic flow is converted into hadrons at freeze-out using the one particle distribution so that energy-momentum tensor is conserved on the hypersurface using relativistic kinetic theory \cite{Cooper:1974mv}. This is supported by the fact that the lattice QCD EoS agrees with that of hadron resonance gas well -- but only approximately -- below the crossover temperature. Also the second-order transport coefficients which appear in causal viscous hydrodynamic equations \cite{Israel:1979wp} are often analytically derived in kinetic theory and the energy density and pressure in the expressions are substituted by the hydrodynamic ones, which partially introduces the effects of interaction to the transport coefficients.

For photons, the parton distribution functions are used in thermal photon emission rates. Oftentimes calculations are performed simply assuming Fermi-Dirac and Bose-Einstein distributions for quarks and gluons, respectively, in the QGP phase, though lattice QCD results seem to deviate from the results implied from those assumptions even at relatively high temperatures \cite{Karsch:2000ps}. Considering the recent status of the photon $v_2$ (and $v_3$) puzzle, the real gas effects should be worth-investigating because the reduction of the effective degrees of freedom at high temperatures could suppress early-time photons with small anisotropy while hadronic photons with larger anisotropy would not be affected much due to the aforementioned agreement of the kinetic theoretical estimations and the lattice QCD data.
There have also been important studies to improve thermal photon models through the phase-space distribution. A few recent examples include viscous corrections to the distribution \cite{Dusling:2008xj,Shen:2014nfa} and semi-QGP effects \cite{Gale:2014dfa}. 

In this paper, I discuss the effects of the interaction in parton phase-space distributions on heavy-ion photons based on the quasi-particle parametrization. There are many variants of the quasi-particle models \cite{Schneider:2001nf, Biro:2001ug, Bluhm:2004qr, Castorina:2005wi, Mattiello:2009fk, Chandra:2011en, Plumari:2011mk}. Here the model is phenomenologically constructed so that it reproduces thermodynamic variables of the lattice QCD equation of state. Then I perform numerical estimations of thermal photon elliptic flow and particle spectra by consistently using the same EoS in a hydrodynamic model and show that modification of the emission rate of the QGP photons leads to visible corrections to the observables. 

In Sec.~\ref{sec2}, the quasi-particle model is developed based on the lattice QCD results. The model for the estimations of thermal photons and background hydrodynamic flow is presented in Sec.~\ref{sec3}. Sec.~\ref{sec4} is devoted for numerical results. Sec.~\ref{sec5} presents conclusions and discussion. The natural units $c = \hbar = k_B = 1$ and the Minkowski metric $g^{\mu \nu} = \mathrm{diag}(+,-,-,-)$ are used in the paper.

\section{Quasi-particle model
}
\label{sec2}
\vspace*{-2mm}

I employ the quasi-particle picture to estimate the effects of in-medium interaction on thermal photon $v_2$, because the photon emission rate is usually given as a functional of distribution functions. Here the quasi-particle model is constructed so that it is compatible with lattice QCD results. The effects of interaction is represented by the (self-)interaction term $W^i_\mathrm{eff} (p,T)$ where $p$ is the four-momentum and $T$ is the temperature. The effective one-particle distribution is
\begin{eqnarray}
f_\mathrm{eff}^i = \frac{1}{\exp{(\omega_i /T)}\pm1} ,
\end{eqnarray}
and the grand-canonical partition function $Z_i$ in a logarithmic form is
\begin{eqnarray}
\ln Z_i = \pm V \int \frac{g_i dp^3}{(2\pi)^3} \ln{\bigg[1\pm \exp{\bigg(- \frac{\omega_i}{T}\bigg) } \bigg]} - \frac{V}{T} \Phi_i (T), \nonumber \\
\end{eqnarray}
where $V$ is the volume, $g_i$ is the degeneracy, $\omega_i = \sqrt{p^2 +m_i^2} + W^i_\mathrm{eff}$ is the effective energy density, and $i$ is the index for particle species, \textit{i.e.}, quarks and gluons. Here the numbers of quarks and antiquarks are assumed to be the same, \textit{i.e.}, the net baryon current is not considered. I consider the light quarks $u$, $d$, and $s$ for the equilibrated quark components. The sign of quantum statistics is $+$ for fermions and $-$ for bosons. $\Phi (T)$ is the background field contribution, which is determined by the thermodynamic consistency condition \cite{Biro:2001ug}
\begin{eqnarray}
\frac{\partial \Phi_i}{\partial T}|_{\mu = 0} = - \int \frac{g_i dp^3}{(2\pi)^3} \frac{\partial \omega_i}{\partial T} f_\mathrm{eff}^i , \label{eq:tcc}
\end{eqnarray}
and the condition $P(T=0) = 0$. This correspond to the bag constant in the bag model \cite{Chodos:1974je} but here it is temperature dependent. 
$W^i_\mathrm{eff}$ is assumed to include the information on interaction and self-energy correction to the hamiltonian and can be regarded as an effective chemical potential as well as thermal correction to the particle mass. 
Energy density and hydrostatic pressure are written as
\begin{eqnarray}
e &=& - \frac{1}{V} \sum_i \frac{\partial \ln Z_i}{\partial \beta} \nonumber \\
&=& \sum_i \int \frac{g_i dp^3}{(2\pi)^3} \bigg(\omega_i - T \frac{\partial \omega^i}{\partial T} \bigg) f^i_\mathrm{eff} + \Phi - T \frac{\partial \Phi}{\partial T} \nonumber \\
&=& \sum_i \int \frac{g_i dp^3}{(2\pi)^3} \omega_i f^i_\mathrm{eff} + \Phi, \label{eq:e} 
\end{eqnarray}
\begin{eqnarray}
P &=& \frac{1}{V} \sum_i T \ln Z_i \nonumber \\
&=& \pm T \sum_i \int \frac{g_i dp^3}{(2\pi)^3} \ln{\bigg[1\pm \exp \bigg(-\frac{\omega_i}{T} \bigg) \bigg]} - \Phi \nonumber \\
&=& \frac{1}{3} \sum_i \int \frac{g_i dp^3}{(2\pi)^3} \mathbf{p} \frac{\partial \omega_i}{\partial \mathbf{p}} f^i_\mathrm{eff} - \Phi , \label{eq:P} 
\end{eqnarray}
where $\beta = 1/T$ and $\Phi = \sum_i \Phi_i$. The last line of Eq.~(\ref{eq:e}) is obtained by using the consistency condition (\ref{eq:tcc}) and that of Eq.~(\ref{eq:P}) by integration by parts. It should be noted that the trace anomaly $\Theta^\mu_\mu = e - 3P$ is no longer vanishing in the massless limit due to the presence of the effective interaction. The entropy density is given through the thermodynamic relation $s = \partial P/\partial T= (e+P)/T$.

Determination of the effective interaction contribution $W^i_\mathrm{eff}$ is generally a non-trivial issue. Here it is constrained with the (2+1)-flavor lattice QCD equation of state \cite{Borsanyi:2013bia} assuming it depends only on the temperature. For simplicity and the lack of additional constraints I assume that $W_\mathrm{eff}$ is common for all partons in this paper to see its qualitative effects.  
It should be noted that a thermodynamically consistent formalism yields the energy density, the pressure, the entropy, and the trace anomaly simultaneously. One can use the entropy density to determine $W_\mathrm{eff}$ as the quantity is independent of $\Phi$, and then determine the background contribution by the relation (\ref{eq:tcc}).

\begin{figure}[tb]
\includegraphics[width=3.0in]{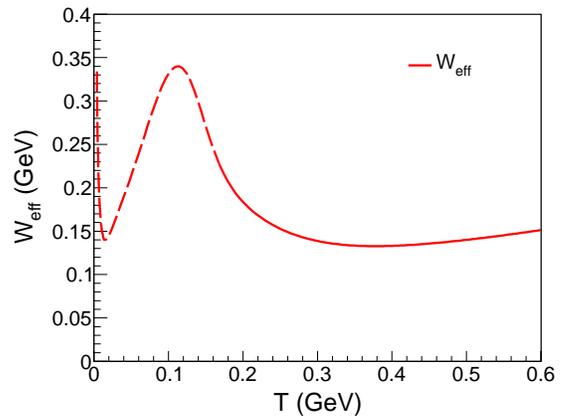}
\caption{(Color online) Lattice data fit of the effective interaction $W_\mathrm{eff}$ as a function of the temperature. Dashed line denotes the region below the crossover. }
\label{fig:weff}
\end{figure}

\begin{figure}[tb]
\includegraphics[width=3.0in]{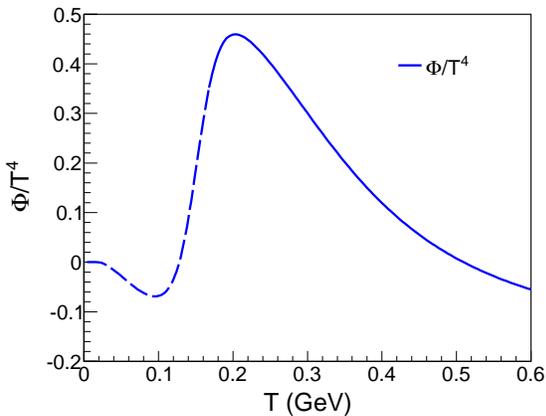}
\caption{(Color online) The background contribution in a dimensionless form $\Phi/T^4$ as a function of the temperature. Dashed line denotes the region below the crossover. }
\label{fig:Phi}
\end{figure}

Figure~\ref{fig:weff} shows the temperature dependence of $W_\mathrm{eff}$ fit to the lattice data. It has a peak structure slightly below $T_c$ because one has to mimic the crossover behavior which does not exist in the free parton gas picture. One can see that the effective fugacity would be non-negligible because it is comparable to the temperature. The corresponding background contribution $\Phi$ in a dimensionless form is shown in Fig.~\ref{fig:Phi}. Compared with the energy density and the pressure, the effective background contribution is small. 
Functional fits to $W_\mathrm{eff}$ and $\Phi$ are presented in Appendix.~\ref{sec:A}.
It should be noted that the parton picture is at most valid above the crossover temperature. 

Although quasi-particle interpretations are commonly employed in hadron physics, I emphasize that the purpose of considering it is to improve the ideal gas distributions often assumed in thermal photon emission rates. The real QCD system is not necessarily a relativistic gas of quasi-particles especially near the crossover. One can also construct similar effective distributions for the hadronic phase but it is not considered here because the difference between the EoS of hadronic resonance gas and that of lattice QCD is indicated to be relatively small as mentioned earlier. 

\section{The model}
\label{sec3}
\vspace*{-2mm}

I develop the numerical model to estimate thermal photon elliptic flow and transverse momentum spectra with and without the effects of interaction corrections using the relativistic hydrodynamic model and the thermal photon emission rates in which the effective distributions obtained in Sec.~\ref{sec2} are embedded. 

\subsection{Thermal photons}

The thermal photon emission rate is affected by the modification of the phase-space distribution. For the demonstrative purpose, the hard photon emissions in Compton scattering and pair annihilation are considered for the QGP phase. I employ the emission rate in Ref.~\cite{Strickland:1994rf} at the leading order in the fugacity expansion by substituting the parton distributions with the effective distributions. The emission rate reads,
\begin{eqnarray}
E\frac{dR}{d^3p} &=& \frac{5 \alpha \alpha_s}{9\pi^2} T^2 e^{-E/T} \bigg\{ \lambda_q \lambda_g \bigg[ \log \bigg( \frac{4ET}{k_c^2} \bigg) + \frac{1}{2} - \gamma \bigg] \nonumber \\
&+& \lambda_q^2 \bigg[ \log \bigg( \frac{4ET}{k_c^2} \bigg) - 1 - \gamma \bigg] \bigg\}. \label{eq:rate} 
\end{eqnarray}
Here $\lambda_q = \lambda_g = e^{-W_\mathrm{eff}/T}$ are the effective fugacities, $\gamma$ is Euler's constant, and $k_c^2 = 2m_\mathrm{th}^2 = g^2T^2/6$ is the infra-red cut-off. 

The hadronic photons are assumed to be unaffected. The emission rate in the hadronic phase is employed from Ref.~\cite{Turbide:2003si}. It should be noted that the photon emission rate near the crossover can be in principle non-trivial \cite{vanHees:2014ida}. Here they are simply interpolated with a hyperbolic function as 
\begin{eqnarray}
E\frac{dR}{d^3p} = c(T) E\frac{dR_\mathrm{lat}}{d^3p} + [1-c(T)] E \frac{dR_\mathrm{had}}{d^3p} ,
\label{eq:approx} 
\end{eqnarray}
where $c(T) = \{1+ \tanh [(T-T_c)/\Delta T]\} /2$ with the connecting temperature $T_c = 0.17$ GeV and the crossover width $\Delta T = 0.017$ GeV. 

\subsection{Hydrodynamic flow}

The flow and the temperature profiles are calculated using the (2+1)-dimensional ideal hydrodynamic model assuming boost-invariance in the space-time rapidity direction \cite{Monnai:2014kqa}. The net baryon number is assumed to be vanishing. The hydrodynamic equation of motion is energy-momentum conservation $\partial_\mu T^{\mu \nu} = 0$ where
\begin{equation}
T^{\mu \nu} = (e+P) u^\mu u^\nu - P g^{\mu \nu} ,
\end{equation}
remains the same for ideal and effective distributions. $u^\mu$ is the flow. The medium property is determined by the EoS, which provides an additional equation for uniquely determining all the thermodynamic variables. Here it is employed from the aforementioned lattice QCD results \cite{Borsanyi:2013bia}, which of course is consistent with the quasi-particle description we use for the estimations of thermal photons. 
The Monte-Carlo version of Glauber model \cite{Miller:2007ri} is used to construct a smoothed initial condition. The numerical code is from Ref.~\cite{Monnai:2014taa}. Here the initial condition is obtained by averaging over events. 

Thermal photon particle spectra can be calculated by integrating the photon emission rate over space-time 
\begin{equation}
\label{eq:spgamma}
\frac{dN^\gamma}{d\phi_p p_Tdp_T dy} = \int dx^4 \frac{dR^\gamma}{d\phi_p p_Tdp_T dy},
\end{equation}
and elliptic flow by taking its second-order Fourier harmonics in the azimuthal angle relative to the reaction plane angle $\Psi$ in momentum space as
\begin{equation}
\label{eq:v2gamma}
v_2^\gamma (p_T,y) = \frac{\int_0 ^{2\pi} d\phi_p \cos (2\phi_p - \Psi) \frac{dN^\gamma}{d\phi_p p_Tdp_T dy}}{\int_0 ^{2\pi} d\phi_p  \frac{dN^\gamma}{d\phi_p p_Tdp_T dy}} .
\end{equation}
Here $p_T$ is the transverse momentum, $y$ is the rapidity, and $\phi_p$ is the angle in momentum space. The local photon energy is estimated as $E=p^\mu u_\mu$.

\section{Numerical results}
\label{sec4}
\vspace*{-2mm}

For demonstrative purposes, Au-Au collisions at $\sqrt{s_{NN}} = 200$~GeV with the impact parameter $b = 6$~fm are considered. The energy density is normalized so that the maximum energy density is $e_\mathrm{max}$ = 30~GeV in the most central collisions at the initial time of hydrodynamic evolution, which is chosen as $\tau_0 = 0.4$~fm/$c$. The contribution of the photon emission above the hadronic freeze-out temperature $T_f = 0.13$ GeV is taken into account.

\subsection{Thermal photon $v_2$}

\begin{figure}[tb]
\includegraphics[width=3.0in]{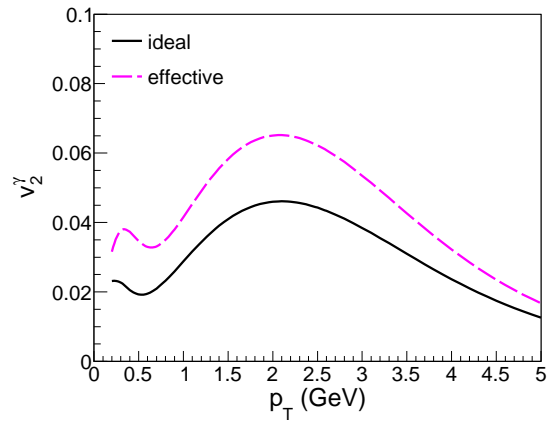}
\caption{(Color online) Thermal photon elliptic flow with effective distributions (dashed line) compared to that with ideal distributions (solid line). }
\label{fig:1}
\end{figure}

The differential thermal photon elliptic flow with the ideal and the effective distributions at mid-rapidity are shown in Fig.~\ref{fig:1} for the transverse momentum window $0.2 < p_T < 5$ GeV. One can see that the effective interaction corrections enhance $v_2^\gamma(p_T)$. This can be understood as follows. The emission of thermal photons are suppressed at early times due to the fact that the effective degrees of freedom in the quasi-particle system is smaller than that in the free gas system when the system is in the QGP phase. The thermal photons emitted at late times have larger elliptic flow because azimuthal anisotropy in the medium develops along with time evolution driven by the pressure gradients. Thus when early-time contribution becomes effectively small, the overall elliptic flow can become large. It should be noted that unlike hadrons, photons are emitted at each space-time point and they all contribute to the final spectra because their interaction with the bulk medium would be weak. 

The effect of the non-ideal gas corrections is consistent with the recently-observed large photon $v_2^\gamma(p_T)$, though the numerical estimations suggest that the effect could be part of the cause for the excessive photon anisotropy but would not be large enough to be the sole reason with the current hydrodynamic parameter sets. 

\subsection{Thermal photon $p_T$ spectra}

\begin{figure}[tb]
\includegraphics[width=3.0in]{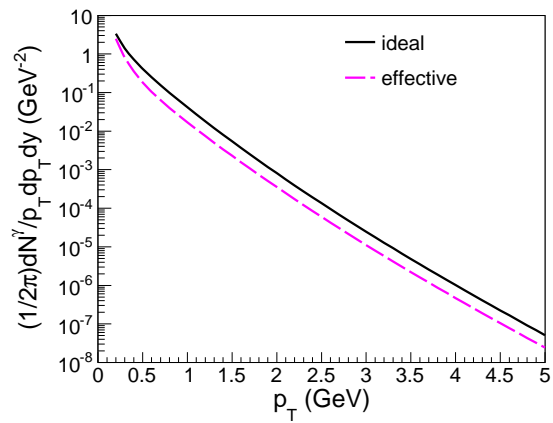}
\caption{(Color online) Thermal photon $p_T$ spectrum with effective distributions (dashed line) compared to that with ideal distributions (solid line). }
\label{fig:2}
\end{figure}

Figure~\ref{fig:2} shows the $p_T$ spectra of thermal photons at mid-rapidity with ideal and effective distributions, respectively. The photon spectrum with the interaction effects is naturally suppressed by the suppression of initial QGP photon emission. This implies that additional photon emission source might be required to explain the spectra when the interaction effects in the phase-space distributions are properly taken into account. Similar discussion can be found, for example, in the system with incomplete quark chemical equilibration \cite{Monnai:2014kqa}. The magnitude of the spectra suppression and $v_2$ enhancement effects are sensitive to the relative ratio of QGP photons to hadronic ones. 

\begin{figure}[tb]
\includegraphics[width=3.0in]{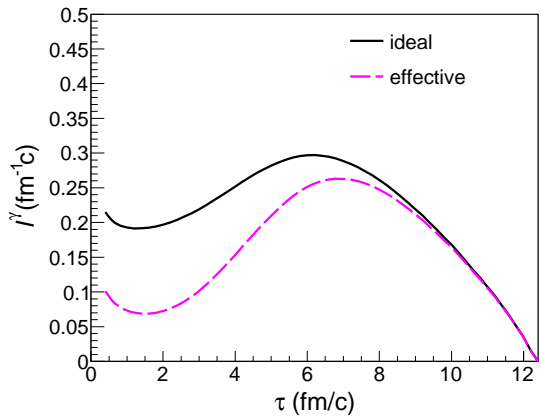}
\caption{(Color online) Spatially-integrated photon emission as a function of proper time with effective distributions (dashed line) compared to that with ideal distributions. }
\label{fig:3}
\end{figure}

The proper time $\tau$ dependence of the photon emission from the medium is also investigated. Here it is defined as the emission rate integrated in the spatial and the transverse momentum directions,
\begin{equation}
\label{eq:I}
I^\gamma (\tau, y) = \int dx^3 \frac{dR^\gamma}{dy} (\tau, \vec{x}, y) .
\end{equation}
The numerical results are shown in Fig.~\ref{fig:3} for the systems with ideal and effective distributions, respectively, at $y = 0$. At the beginning photon production is reduced along with time evolution because the temperature decreases rapidly due to the longitudinal expansion. The in-medium corrections suppress the emission by a factor of $\sim e^{-2W_\mathrm{eff}/T}$, which becomes larger as it nears the pseudo-critical temperature. On the other hand, hadronic photon contributions become larger as the system cools down and consequently the difference between the two systems disappears at later times.

\section{Discussion and Conclusions}
\label{sec5}
\vspace*{-2mm}

Effects of the interaction corrections in parton distributions on heavy-ion thermal photons are investigated. This allows one to preserve the consistency between the hydrodynamic evolution and the photon emission rate. The quasi-particle model is introduced and constrained so that it reproduces the thermodynamic variables from the lattice QCD estimations. Thermal photon emission is estimated in numerical simulations, and it is found that the suppression of the QGP photons in early stages leads to enhancement of thermal photon $v_2$. $p_T$ spectrum of thermal photons, on the other hand, is suppressed by the corrections. The results imply that one has to take the modifications of phase-space distributions into account for the quantitative description of heavy-ion photon spectra and flow harmonics. Since the mechanism of early-time photon suppression can also lead to the enhancement of thermal photon $v_3$, it would be important to perform event-by-event analyses for calculating higher-order flow harmonics.

One should be careful before quantitative discussion that prompt photons, which will reduce the overall anisotropy in direct photons, are not included here and that only hard photons above $k_c^2$ are considered for the QGP photons. Also though the corrections to ideal gas distributions would be a necessarily ingredient in the heavy-ion modeling and the magnitude of $v_2$ enhancement is visible in the model calculations, it would not be enough to fully account for the large direct photon $v_2$ observed in the experiments especially with prompt photon contributions. 

There would be several interesting applications of the quasi-particle model to heavy-ion phenomenology. Firstly, the formalism can provide a model equation of state for chemical non-equilibrated quark-gluon systems which is compatible with the lattice QCD EoS in the equilibrium limit, which is useful in studying the effects of quark chemical equilibration in hydrodynamic systems \cite{Monnai:2014kqa}. Secondly, it would be interesting to apply the effective distributions for the derivation of causal dissipative hydrodynamic equations because some formalism, such as Israel-Stewart theory \cite{Israel:1979wp}, are explicitly dependent on the equilibrium distribution. The transport coefficients can then be fully consistent with hydrodynamic EoS. Thirdly, this may also allow one to formulate an anisotropic hydrodynamic formalism which is compatible with the lattice EoS by using the effective distribution instead of the ideal gas distribution.

\begin{acknowledgments}
The author would like to thank M.~Asakawa, M.~Kitazawa, K.~Morita, A.~Ohnishi, and B.~Schenke for insightful comments.
The work of A.M. is supported by RIKEN Special Postdoctoral Researcher program. Some of the results are calculated using RIKEN Integrated Cluster of Clusters (RICC).
\end{acknowledgments}

\appendix

\section{Functional Fits for the Effective Distributions}
\label{sec:A}

In this appendix, the polynomial fits to the in-medium correction parameters in the parton phase-space distributions are presented. The contribution of gluons and up, down, and strange quarks are considered.
The fit to the effective interaction energy $W_\mathrm{eff}$ for $0.15 < T < 0.6$ GeV is given as
\begin{eqnarray}
W_\mathrm{eff} &=& a_2 t^2 + a_1 t + a_0 + \frac{a_{-1}}{t} + \frac{a_{-2}}{t^2} + \frac{a_{-3}}{t^3} + \frac{a_{-4}}{t^4} ,\nonumber \\
\end{eqnarray}
where $a_2 = -1.040$, $a_1 = 2.419$, $a_0 = -1.879$, $a_{-1} = 0.8181$, $a_{-2} = -0.1796$, $a_{-3} = 0.02072$, and $a_{-4} = -0.0009111$. 
Here $t = T$ GeV/1 GeV is defined as the dimensionless temperature. 

Likewise, the background contribution $\Phi$ would be given by the following functional fit:
\begin{eqnarray}
\Phi &=& b_3 t^3 + b_2 t^2 + b_1 t + b_0,
\end{eqnarray}
where $b_3 = -0.3303$, $b_2 = 0.2424$, $b_1 = -0.04152$, and $b_0 = 0.001981$ for the aforementioned temperature range.

\bibliography{basename of .bib file}

\begin{thebibliography}{99}

\bibitem{RHIC:summary}
  K.~Adcox {\it et al.}  [PHENIX Collaboration],
  Nucl.\ Phys.\ A {\bf 757}, 184 (2005);
  J.~Adams {\it et al.}  [STAR Collaboration],
  Nucl.\ Phys.\ A {\bf 757}, 102 (2005);
  B.~B.~Back {\it et al.} [PHOBOS Collaboration],
  Nucl.\ Phys.\ A {\bf 757}, 28 (2005);
  I.~Arsene {\it et al.}  [BRAHMS Collaboration],
  Nucl.\ Phys.\ A {\bf 757}, 1 (2005).
  
\bibitem{LHC:v2} 
  K.~Aamodt {\it et al.}  [The ALICE Collaboration],
  Phys.\ Rev.\ Lett.\  {\bf 105}, 252302 (2010);
  G.~Aad {\it et al.}  [ATLAS Collaboration],
  Phys.\ Lett.\ B {\bf 707}, 330 (2012);
  S.~Chatrchyan {\it et al.}  [CMS Collaboration],
  Eur.\ Phys.\ J.\ C {\bf 72}, 2012 (2012).

\bibitem{Schenke:2010rr} 
  B.~Schenke, S.~Jeon and C.~Gale,
  Phys.\ Rev.\ Lett.\  {\bf 106}, 042301 (2011).

\bibitem{Ollitrault:1992bk} 
  J.~-Y.~Ollitrault,
  Phys.\ Rev.\ D {\bf 46}, 229 (1992).
  
\bibitem{Poskanzer:1998yz} 
  A.~M.~Poskanzer and S.~A.~Voloshin,
  Phys.\ Rev.\ C {\bf 58}, 1671 (1998).
  
\bibitem{Adare:2011zr} 
  A.~Adare {\it et al.}  [PHENIX Collaboration],
  Phys.\ Rev.\ Lett.\  {\bf 109}, 122302 (2012).
  
\bibitem{Lohner:2012ct} 
  D.~Lohner [ALICE Collaboration],
  J.\ Phys.\ Conf.\ Ser.\  {\bf 446}, 012028 (2013).
  
\bibitem{Mizuno:2014via} 
  S.~Mizuno [PHENIX Collaboration],
  Nucl.\ Phys.\ A {\bf 931}, 686 (2014).
  
\bibitem{Chatterjee:2005de} 
  R.~Chatterjee, E.~S.~Frodermann, U.~W.~Heinz and D.~K.~Srivastava,
  Phys.\ Rev.\ Lett.\  {\bf 96}, 202302 (2006).
  
\bibitem{Chatterjee:2008tp} 
  R.~Chatterjee and D.~K.~Srivastava,
  Phys.\ Rev.\ C {\bf 79}, 021901 (2009).
  
\bibitem{Chatterjee:2011dw} 
  R.~Chatterjee, H.~Holopainen, T.~Renk and K.~J.~Eskola,
  Phys.\ Rev.\ C {\bf 83}, 054908 (2011).
  
\bibitem{Holopainen:2011pd} 
  H.~Holopainen, S.~Rasanen and K.~J.~Eskola,
  Phys.\ Rev.\ C {\bf 84}, 064903 (2011).
  
\bibitem{Chatterjee:2013naa} 
  R.~Chatterjee, H.~Holopainen, I.~Helenius, T.~Renk and K.~J.~Eskola,
  Phys.\ Rev.\ C {\bf 88}, 034901 (2013).
  
\bibitem{Chatterjee:2014nta} 
  R.~Chatterjee, D.~K.~Srivastava and T.~Renk,
  arXiv:1401.7464 [hep-ph].
  
\bibitem{Liu:2009kta} 
  F.~-M.~Liu, T.~Hirano, K.~Werner and Y.~Zhu,
  Phys.\ Rev.\ C {\bf 80}, 034905 (2009).
  
\bibitem{vanHees:2011vb} 
  H.~van Hees, C.~Gale and R.~Rapp,
  Phys.\ Rev.\ C {\bf 84}, 054906 (2011).
  
\bibitem{Dion:2011pp} 
  M.~Dion, J.~-F.~Paquet, B.~Schenke, C.~Young, S.~Jeon and C.~Gale,
  Phys.\ Rev.\ C {\bf 84}, 064901 (2011).
  
\bibitem{Basar:2012bp} 
  G.~Basar, D.~Kharzeev, and V.~Skokov,
  Phys.\ Rev.\ Lett.\  {\bf 109}, 202303 (2012);
%
  G.~Basar, D.~E.~Kharzeev and E.~V.~Shuryak,
  Phys.\ Rev.\ C {\bf 90}, 014905 (2014).
  
\bibitem{Bzdak:2012fr} 
  A.~Bzdak and V.~Skokov,
  Phys.\ Rev.\ Lett.\  {\bf 110}, 192301 (2013).
  
\bibitem{Goloviznin:2012dy} 
  V.~V.~Goloviznin, A.~M.~Snigirev and G.~M.~Zinovjev,
  JETP Lett.\  {\bf 98}, 61 (2013).
  
\bibitem{Hattori:2012je} 
  K.~Hattori and K.~Itakura,
  Annals Phys.\  {\bf 330}, 23 (2013);
  %
  Annals Phys.\  {\bf 334}, 58 (2013).
  
\bibitem{Linnyk:2013hta} 
  O.~Linnyk, V.~P.~Konchakovski, W.~Cassing and E.~L.~Bratkovskaya,
  Phys.\ Rev.\ C {\bf 88}, 034904 (2013);
%
  O.~Linnyk, W.~Cassing and E.~L.~Bratkovskaya,
  Phys.\ Rev.\ C {\bf 89}, 034908 (2014).
  
\bibitem{Liu:2012ax} 
  F.~-M.~Liu and S.~-X.~Liu,
  Phys.\ Rev.\ C {\bf 89}, 034906 (2014).
  
\bibitem{Muller:2013ila} 
  B.~M\"{u}ller, S.~-Y.~Wu and D.~-L.~Yang,
  Phys.\ Rev.\ D {\bf 89}, 026013 (2014).
  
\bibitem{Monnai:2014kqa} 
  A.~Monnai,
  Phys.\ Rev.\ C {\bf 90}, 021901 (2014);
  arXiv:1410.8621 [nucl-th];
  arXiv:1412.7781 [nucl-th].
  
\bibitem{McLerran:2014hza} 
  L.~McLerran and B.~Schenke,
  Nucl.\ Phys.\ A {\bf 929}, 71 (2014).
  
\bibitem{Monnai:2014taa} 
  A.~Monnai,
  arXiv:1408.1410 [nucl-th].
  
\bibitem{vanHees:2014ida} 
  H.~van Hees, M.~He and R.~Rapp,
  Nucl.\ Phys.\ A {\bf 933}, 256 (2015).
  
\bibitem{Shen:2013cca} 
  C.~Shen, U.~W.~Heinz, J.~F.~Paquet, I.~Kozlov and C.~Gale,
  Phys.\ Rev.\ C {\bf 91}, 024908 (2015).
  
\bibitem{Gale:2014dfa} 
  C.~Gale {\it et al.}, 
  Phys.\ Rev.\ Lett.\  {\bf 114}, 072301 (2015).
  
\bibitem{Biro:2015iua} 
  T.~S.~Bir\'{o}, M.~Horv\'{a}th and Z.~Schram,
  arXiv:1503.06628 [hep-ph].
  
\bibitem{Cooper:1974mv} 
  F.~Cooper and G.~Frye,
  Phys.\ Rev.\ D {\bf 10}, 186 (1974).
  
\bibitem{Israel:1979wp} 
  W.~Israel and J.~M.~Stewart,
  Annals Phys.\  {\bf 118}, 341 (1979).
  
\bibitem{Karsch:2000ps} 
  F.~Karsch, E.~Laermann and A.~Peikert,
  Phys.\ Lett.\ B {\bf 478}, 447 (2000);
  A.~Bazavov {\it et al.}  [HotQCD Collaboration],
  Phys.\ Rev.\ D {\bf 90}, 094503 (2014).
  
\bibitem{Dusling:2008xj} 
  K.~Dusling and S.~Lin,
  Nucl.\ Phys.\ A {\bf 809}, 246 (2008);
%
  K.~Dusling,
  Nucl.\ Phys.\ A {\bf 839}, 70 (2010).
  
\bibitem{Shen:2014nfa} 
  C.~Shen, J.~F.~Paquet, U.~Heinz and C.~Gale,
  Phys.\ Rev.\ C {\bf 91}, 014908 (2015).
  
\bibitem{Schneider:2001nf} 
  R.~A.~Schneider and W.~Weise,
  Phys.\ Rev.\ C {\bf 64}, 055201 (2001);
  T.~Renk, R.~A.~Schneider and W.~Weise,
  Phys.\ Rev.\ C {\bf 66}, 014902 (2002)
  
\bibitem{Biro:2001ug} 
  T.~S.~Biro, A.~A.~Shanenko and V.~D.~Toneev,
  Phys.\ Atom.\ Nucl.\  {\bf 66}, 982 (2003).
  
\bibitem{Bluhm:2004qr} 
  M.~Bluhm, B.~Kampfer and G.~Soff,
  J.\ Phys.\ G {\bf 31}, S1151 (2005);
  M.~Bluhm, B.~Kampfer, R.~Schulze and D.~Seipt,
  Eur.\ Phys.\ J.\ C {\bf 49}, 205 (2007).
  
\bibitem{Castorina:2005wi} 
  P.~Castorina and M.~Mannarelli,
  Phys.\ Lett.\ B {\bf 644}, 336 (2007).
  
\bibitem{Mattiello:2009fk} 
  S.~Mattiello and W.~Cassing,
  J.\ Phys.\ G {\bf 36}, 125003 (2009).
  
\bibitem{Chandra:2011en} 
  V.~Chandra and V.~Ravishankar,
  Phys.\ Rev.\ D {\bf 84}, 074013 (2011).
  
\bibitem{Plumari:2011mk}
  S.~Plumari, W.~M.~Alberico, V.~Greco and C.~Ratti,
  Phys.\ Rev.\ D {\bf 84}, 094004 (2011).
  
\bibitem{Chodos:1974je} 
  A.~Chodos, R.~L.~Jaffe, K.~Johnson, C.~B.~Thorn and V.~F.~Weisskopf,
  Phys.\ Rev.\ D {\bf 9}, 3471 (1974).
 
\bibitem{Borsanyi:2013bia} 
  S.~Borsanyi, Z.~Fodor, C.~Hoelbling, S.~D.~Katz, S.~Krieg and K.~K.~Szabo,
  Phys.\ Lett.\ B {\bf 730}, 99 (2014).
  
\bibitem{Strickland:1994rf} 
  M.~Strickland,
  Phys.\ Lett.\ B {\bf 331}, 245 (1994);
%
  D.~K.~Srivastava, M.~G.~Mustafa and B.~M\"{u}ller,
  Phys.\ Rev.\ C {\bf 56}, 1064 (1997).
  
\bibitem{Turbide:2003si} 
  S.~Turbide, R.~Rapp and C.~Gale,
  Phys.\ Rev.\ C {\bf 69}, 014903 (2004);
%
  F.~Arleo, P.~Aurenche, F.~W.~Bopp, I.~Dadic, G.~David, H.~Delagrange, D.~G.~d'Enterria and K.~J.~Eskola {\it et al.},
  hep-ph/0311131.

\bibitem{Miller:2007ri} 
  M.~L.~Miller, K.~Reygers, S.~J.~Sanders and P.~Steinberg,
  Ann.\ Rev.\ Nucl.\ Part.\ Sci.\  {\bf 57}, 205 (2007).

\end{thebibliography}

\end{document}